\documentclass[fleqn,twoside,11pt]{article}
\usepackage{espcrc2}
\usepackage{hyperref}

\usepackage{graphicx}
\usepackage[figuresright]{rotating}
\usepackage{amsmath}
\usepackage{textgreek}
\usepackage{footnote}
\usepackage{geometry}
\def\gsim{\raise0.3ex\hbox{$>$\kern-0.75em\raise-1.1ex\hbox{$\sim$}}}
\def\lsim{\raise0.3ex\hbox{$<$\kern-0.75em\raise-1.1ex\hbox{$\sim$}}}

\title{{\bf CMB {\it B}-modes, spinorial space-time and Pre-Big Bang (II)} \vspace{2mm}}

\author{{\bf Luis Gonzalez-Mestres}\thanks{luis.gonzalez-mestres@megatrend.edu.rs at the Megatrend Cosmology Laboratory \newline Luis.Gonzalez-Mestres@univ-savoie.fr at the Universit\'{e} de Savoie \newline lgmsci@yahoo.fr, personal e-mail} \vspace{2mm}
 \\ {\bf Megatrend Cosmology Laboratory, Megatrend University, Belgrade and Paris \\ Goce Delceva 8, 11070 Novi Beograd, Serbia} \vspace{2mm}}
             
\begin{document}

\begin{abstract}
{\bf The BICEP2 collaboration reported recently a {\it B}-mode polarization of the cosmic microwave background (CMB) radiation inconsistent with the null hypothesis at a significance of $>$ 5 $\sigma $. This result has been often interpreted as a signature of primordial gravitational waves from cosmic inflation, even if actually polarized dust emission may be at the origin of such a signal. Even assuming that part of this CMB $B$-mode polarization really corresponds to the early Universe dynamics, its interpretation in terms of inflation and primordial gravitational waves is not the only possible one. Alternative cosmologies such as pre-Big Bang patterns and the spinorial space-time (SST) we introduced in 1996-97 can naturally account for such CMB $B$-modes. In particular, the SST automatically generates a privileged space direction (PSD) whose existence may have been confirmed by Planck data. If such a PSD exists, it seems normal to infer that vector perturbations have been present in the early Universe leading to CMB $B$-modes in suitable cosmological patterns. Inflation would not be required to explain the BICEP2 result assuming it really contains a primordial signal. More generally, pre-Big Bang cosmologies can also generate gravitational waves in the early Universe without any need for cosmic inflation. We further discuss here possible alternatives to the inflationary interpretation of a primordial $B$-mode polarization of cosmic microwave background radiation.}  
\vspace{1pc}
\end{abstract}

\maketitle

\section{Introduction}
BICEP2 data \cite{BICEP2-I,BICEP2-II} may have provided a signature of CMB $B$-modes from the early universe. However, this is not yet certain as the observed signal could actually be due to galactic dust effects \cite{Liu,PlanckXIX} even if the experimental program looks promising in all cases \cite{Mortonson,Flauger}. 

Furthermore, if the experimental and phenomenological uncertainty remains, the situation is not different from a theoretical point of view, as already pointed out in \cite{Gonzalez-Mestres2014}.

Assuming that some of the recent BICEP2 data really correspond to a primordial $B$-mode polarization of cosmic microwave background radiation, the theoretical interpretation of such a signal would not yet be obvious. Alternative cosmologies must be seriously considered \cite{Gonzalez-Mestres2014,Gonzalez-MestresCrete2013} and can even work more naturally than the standard Big Bang with cosmic inflation to explain the observed effect. 

The BICEP2 result is often presented as a strong direct evidence for cosmic inflation and primordial gravitational waves. It is claimed that the $B$-modes of CMB cannot be generated primordially by scalar (density) perturbations and that only gravitational waves (tensor perturbations) originating from the inflationary expansion of the Universe can produce them. Possible vector perturbations (vorticity), that can in principle naturally generate such $B$-modes, are not really considered within the conventional cosmological framework. But this kind of analysis applies only to cosmologies based on the standard Big Bang approach where inflation is a crucial ingredient.

The situation can be radically different in reasonable alternative cosmologies. Pre-Big Bang models \cite{Gonzalez-MestresCrete1,Gonzalez-MestresPlanckSST} do not in general require an inflationary scenario and can efficiently produce primordial CMB $B$-modes through vector perturbations \cite{Gonzalez-Mestres2014}. The spinorial space-time (SST) we suggested in 1996-97 \cite{Gonzalez-Mestres1996,Gonzalez-Mestres1997a} automatically generates \cite{Gonzalez-MestresPrivSpDir,Gonzalez-MestresSSTFriedmann} a privileged space direction (PSD) for each comoving observer. Then, the existence of CMB $B$-modes is a natural consequence of this space anisotropy of geometric and cosmic origin.

Alternative cosmologies, including pre-Big Bang, are not "exotic" and have not been excluded by data. Physics beyond the Planck scale can be a natural extension of standard theories if quantum mechanics ceases to hold or undergoes modifications at very high energies and very small distances \cite{Gonzalez-MestresCrete1,Gonzalez-MestresCrete2013bis}. Similarly, the effective space-time structure can depend on the energy scale \cite{Gonzalez-Mestres1995,Gonzalez-MestresCRIS2010}. In the case of SST, the existence of a privileged space direction for each comoving observer, already compatible with WMAP data \cite{WMAP}, may have been confirmed by more recent Planck \cite{Planck} results \cite{PlanckAniso}. The PSD combined with parity violation can potentially explain the observed CMB anisotropy \cite{Gonzalez-MestresPlanckSST,Gonzalez-MestresPrivSpDir}. Pre-Big Bang models can naturally solve the horizon problem \cite{Gonzalez-Mestres1995} and provide sensible alternatives to the inflationary description of the formation of conventional matter structure in our Universe. They can also generate primordial gravitational waves without the standard cosmic inflation, as shown in an approach based on an initial gravitational instanton at cosmic time $t$ = 0 \cite{Bogd1,Bogd2}.  

In this note, we further develop the analysis of \cite{Gonzalez-Mestres2014} on possible alternatives to the inflationary interpretation of BICEP2 results assuming the $B$-modes of CMB really correspond, at least partially, to a signal from the early Universe.

\section{The spinorial space-time (SST)}

It is well known that fermions do not feel space exactly in the same way as bosons and macroscopic objects. In particular, a 360 degrees rotation changes the sign of a spin-1/2 wave function. To explore all possible consequences of this property of particles with half-integer spin, we introduced \cite{Gonzalez-Mestres1996,Gonzalez-Mestres1997a} a spinorial SU(2) space-time with two complex coordinates replacing the four standard real ones. The properties of the SST, including its possible cosmological implications, have been reminded and further studied in \cite{Gonzalez-MestresCrete1,Gonzalez-MestresCRIS2010} and in \cite{Gonzalez-MestresPrivSpDir,Gonzalez-MestresSSTFriedmann}. 

\subsection{SST basic structure}
\vskip2mm
It turns out that the use of the SST instead of the conventional real space-time can have important and connected implications for both the internal properties of standard elementary particles and the very large scale structure of the Universe. The two domains are actually related through pre-Big Bang evolution, where the ultimate structure of matter and space-time is expected to play a leading role and dominate the overall dynamics of the primordial Universe.  

To a SU(2) spinor $\xi $ describing cosmic space-time coordinates (two complex variables), it is possible to associate a positive SU(2) scalar $\mid \xi \mid ^2$ = $\xi ^\dagger \xi $ (the dagger stands for hermitic conjugate). A definition of cosmic time (age of the Universe) can then be $t$ = $\mid \xi \mid$ with an associated space given by the $S^3$ hypersphere $\mid \xi \mid$ = $t$. Other definitions of $t$ in terms of $\mid \xi \mid$ (f.i. $t$ = $\mid \xi \mid ^2$) lead to similar cosmological results as long as a single-valued function is used. 

With the definition $t$ = $\mid \xi \mid$, if $\xi _0$ is the observer position on the $\mid \xi \mid $ = $t_0$ hypersphere, space translations inside this hypersphere are described by SU(2) transformations acting on the spinor space, i.e. $\xi ~=~ U ~\xi _0$ with:
\begin{equation}
U~=~exp~(i/2~~t_0^{-1}~{\vec{\mathbf \sigma }}.{\vec{\mathbf x}})~ \equiv U ({\vec{\mathbf x}})
\end{equation}
where ${\vec{\mathbf \sigma }}$ is the vector formed by the usual Pauli matrices and the vector ${\vec{\mathbf x}}$ the spatial position (in time units, at that stage) of $\xi $ with respect to $\xi _0$ at constant time $t_0$. 

The origin of cosmic time, associated to the beginning of the Universe, is naturally given by the point $\xi$ = 0 where the initial space is contracted to a single point. One then has an expanding universe where cosmological comoving frames correspond to straight lines going through the time origin $\xi $ = $0$. The SST geometry naturally suggests the existence of a local privileged rest frame for each comoving observer, which is compatible with existing cosmological observations. 

As already pointed out in \cite{Gonzalez-Mestres1996,Gonzalez-Mestres1997a}, an attempt to associate to the cosmic spinor $\xi $ real cosmic space coordinates ${\vec{\mathbf x_c}}$ writing ${\vec{\mathbf x_c}}$ = $\xi ^\dagger {\vec{\mathbf \sigma }} \xi $ does not really lead to such coordinates. Instead, one gets $\mid \xi \mid ^2$ times a unit vector defining the local PSD. The standard space coordinates can only be defined from an origin $\xi _0$ at the same cosmic time $t$, as in (1). This situation clearly illustrates the potential limitations of general relativity and standard cosmology. Rather than an intrinsic fundamental property of space and time, conventional relativity would most likely be a low-energy symmetry of standard matter similar to the well-known effective Lorentz-like symmetry of the kinematics of low-momentum phonons in a lattice \cite{Gonzalez-Mestres1997b,Gonzalez-Mestres1995bis} where the speed of sound plays the role of the critical speed. The speed of light would then be the critical speed of a family of vacuum excitations (the standard particles) not directly related to an intrinsic space-time geometry.   

Space rotations with respect to a fixed point $\xi _0$ are given by SU(2) transformations acting on the spatial position vector ${\vec{\mathbf x}}$ defined by (1). A standard spatial rotation around $\xi _0 $ is now given by a SU(2) element $U({\vec{\mathbf y}})$ turning $U({\vec{\mathbf x}})$ into $U({\vec {\mathbf y}}) ~ U({\vec{\mathbf x}}) ~ U({\vec{\mathbf y}})^\dagger $. The vector $\vec{\mathbf y}$, related to $U({\vec{\mathbf y}})$ in a similar way to (1), provides the rotation axis and angle. If a spin-1/2 particle is present at the position ${\vec{\mathbf x}}$ with an associated spinor $\xi _p$ describing its spin, then $\xi _p$ transforms into $\xi _p'$ = $U({\vec{\mathbf y}})$ $\xi _p$.

\subsection{Some direct consequences}
\vskip2mm
It can be readily checked \cite{Gonzalez-MestresCrete1,Gonzalez-MestresCRIS2010} that the SST automatically generates two basic cosmological phenomena in a purely geometric way :

i) The standard relation between relative velocities and distances at cosmic scale, with a ratio $H$ (velocity/distance) equal to the inverse of the age of the Universe ($H ~ = ~t^{-1}$).  

ii) As previously stressed, a privileged space direction for each comoving observer.

Furthermore, space translations form a (non-abelian) compact group, contrary to standard space-time geometry.

The PSD associated to the space-time point $\xi $ is defined by the linear combination of sigma matrices (with real coefficients) that leaves $\xi $ invariant. The space-time points on the trajectory generated by this sigma-like matrix satisfy the relation $\xi ' ~ = ~ exp ~(i \phi )~ \xi $ where $\phi $ is real and $ exp ~(i \phi )$ is a complex phase factor. This definition of the PSD is stable under SU(2) transformations and comoving time evolution. 

The existence of the PSD is a cosmological property specific to the spinorial structure of the cosmic space-time as "seen" from the cosmic origin $\xi $ = 0 ($t$ = 0). The PSD is not apparent in the space-time geometry when standard space coordinates (the above ${\vec{\mathbf x}}$) are used, as these coordinates belong to a vector representation of SU(2) and SO(3). Thus, conventional cosmology based on the usual real space-time cannot in principle account for the PSD in a simple way. We expect bosons and usual macroscopic objects to be less directly concerned by PSD effects than the elementary fermions, the possible ultimate constituents of matter and the very large scale structure of the Universe. 

Contrary to the standard isotropic description of the early Universe, where only $E$-modes associated to gradients are assumed to be present in the CMB except for the $B$-modes due to inflationary gravitational waves, a cosmology based on the spinorial space-time naturally leads to $B$-modes generated by rotations around the privileged space direction and vector products by this direction. Then, cosmic inflation and primordial gravitational waves are no longer necessary to account for the primordial CMB $B$-modes that BICEP2 has possibly observed. On the contrary, such a result, together with recent Planck data, may have provided a signature of SST geometry or of some other unconventional structure beyond the standard space-time and cosmology.

\subsection{SST, conservation laws, causality}
\vskip2mm
The existence of a PSD implies a potential violation of rotation invariance in Particle Physics that may invalidate the standard conservation law for angular momentum in phenomena sensitive to the PSD. However, such an effect can be very difficult to detect in Particle Physics experiments, as orbital angular momentum is defined using position and momentum operators that are vector representations of the space symmetry group. The internal structure of fermions would be sensitive to the PSD and potentially lead to some observable signatures.

As the time-dependent global size of the Universe is part of the fundamental space-time geometry, one can consider that energy conservation does no longer follow as an exact basic law of Physics. Even if we expect the possible effects of energy non conservation due to the Universe expansion to be too small to be detected in laboratory experiments, the possible evolution of vacuum structure and particle properties at cosmological scales must be carefully explored.  

When using the SST geometry for conventional particles, it seems normal to describe the internal structure of standard elementary fermions (quarks and leptons) through a spinorial wave function defined in a local SST whose origin lies at the particle space-time position. Then, for a comoving particle at $\xi _0$, the local spinorial coordinates of a point $\xi $ would be given by the spinor $\xi _L$ = $\xi $ - $\xi _0$. Considering a wave function of the type $\Psi (\xi _L)$ to describe the lepton and quark internal structure \cite{Gonzalez-MestresCrete1,Gonzalez-MestresCRIS2010} provides an unconventional alternative to standard causality at very small distance and local time scales, as the values of $\xi $ thus considered do not in general correspond to the same value of the cosmic time as $\xi _0$. At such scales, the notion of time itself should be renconsidered. 

Assuming the internal wave function of a standard "elementary" particle to be a SU(2) eigenstate, the allowed spin (spinorial angular momentum) values would be multiples of 1/2, including 0, 1/2, 1, 3/2 and 2 but also possibly higher spins contrary to standard assumptions. 

All particles of standard physics can thus be generated by a spinorial wave function, and the existence of "elementary" spin-3/2 particles seems natural in such a pattern. As the Poincar\'{e} group is no longer an exact symmetry in such an approach, an alternative to supersymmetry involving both space-time and internal symmetries may thus emerge as a new (approximate and broken) symmetry escaping standard theorems \cite{Gonzalez-MestresCrete1,Gonzalez-MestresCRIS2010}. The subject clearly requires further exploration, including the experimental search for signatures of such "elementary" (similar to quarks and leptons) spin-3/2 particles and of possible spin-2 particles other than the graviton. 

Similarly, the possible existence of "elementary" particles with spin larger than 2 cannot be excluded in such a pattern and deserves close theoretical and experimental investigation. High spin elementary fields were already considered in a different approach \cite{Vasiliev1,Vasiliev2}. An alternative to standard quantum field theory (SQFT) where the basic vacuum structure is not dominated by the usual field condensates and zero modes has been suggested in \cite{Gonzalez-Mestres2009,Gonzalez-MestresIU} and in \cite{Gonzalez-MestresCrete1,Gonzalez-MestresCRIS2010}.  

\section{Pre-Big Bang, SST, PSD, superbradyons, $B$-modes...}

In this and previous articles, we always consider pre-Big Bang scenarios that are not based on mere extrapolations from standard dynamics (including strings) to higher energies and lower distance scales. Our basic assumption about pre-Big Bang is that really new physics leads the Universe evolution at distance and time scales smaller than the Planck scale, and that the standard principles of Physics (relativity, quantum mechanics...) cease to be valid at these scales or even at larger scales \cite{Gonzalez-MestresPreBB1,Gonzalez-MestresPreBB2}. New ultimate constituents of matter and a new space-time geometry can then dominate this unconventional primordial phase of the history of the Universe. Conventional gravity is not necessarily the appropriate framework to understand the ultimate origin of space and time.

In spite of its already important implications, the above described SST does not yet incorporate space units, standard matter or even a definite vacuum structure. The size of the SST-based universe can be much larger than that of the conventional one. It may even happen that standard matter occupies only a small part of the SST, or that its nucleation has occurred in many independent regions. From a dynamical point of view, it seems normal to assume that the SST geometry is somehow in quasi-equilibrium with an underlying physical vacuum, even if the SST structure and the evolution of the Universe (the time-dependent radius) reflect by themselves the existence of dominant cosmic forces leading to this evolution in time. As previously stressed, the notion of time itself deserves further thought \cite{Merali}.  

If the vacuum is made of a fundamental matter or pre-matter different from standard matter and of which the conventional "elementary" particles are actually composite, the speed of light is not expected to be a fundamental critical speed. The ultimate matter constituents can have a critical speed much faster than that of light just as the speed of light is much faster than that of sound \cite{Gonzalez-Mestres1995,Gonzalez-Mestres1997b}. Then, it is not excluded that the ultimate fundamental objects (such as superbradyons \cite{Gonzalez-Mestres1995bis}) exist in our Universe as free superluminal particles. They can be remnants from the early Universe \cite{Gonzalez-Mestres1996,Gonzalez-MestresCRIS2010} and part of the dark matter \cite{Gonzalez-MestresPreBB1,Gonzalez-Mestres2009}.

\subsection{A new Friedmann-like equation}
\vskip2mm
As emphasized in \cite{Gonzalez-MestresCrete2013,Gonzalez-MestresPlanckSST}, the SST leads to a new approach to the role of space curvature in cosmology and to a new structure of Friedmann-like equations. In particular, the leading contribution to the square of the Lundmark-Lema\^itre-Hubble constant \cite{Gonzalez-MestresCrete1} $H$ comes from a curvature term equal to $t^{-2}$ whose sign does not depend on the space curvature felt by standard matter. In \cite{Gonzalez-MestresCrete2013}, the following Friedmann-like relation was considered :
\begin{equation}
H^2 ~=~ 8 \pi ~G ~\rho /3 ~-~ k ~R^{-2} ~c^{2}~+~t^{-2}~+~ \textKappa ~+~\Lambda ~c^2/3
\end{equation}  
where $\rho $ is the energy density associated to standard matter, $c$ the speed of light, $k~ R^{-2}$ the curvature parameter, $R$ the present curvature distance scale of the Universe (the curvature radius, and possibly the radius of the Universe, for $k$ = 1) and $\Lambda $ a possible new version of the cosmological constant decreasing like the matter density as the Universe expands. The new term $t^{-2}$, of cosmic geometric origin as suggested by the SST structure, dominates the large scale expansion of the Universe. $\textKappa $ is a correction term accounting in particular for:

- a possible small difference between the comoving frames of standard cosmology and those (pre-existing) obtained from the underlying SST cosmic geometry;

- a reaction of the nucleated standard matter to the pre-existing expansion of the Universe led by the SST geometry \cite{Gonzalez-MestresCrete1,Gonzalez-MestresPlanckSST};

- vacuum inhomogeneities at cosmic scale and other non-standard effects.

A further modification of (2) accounting for phenomena related to the local privileged space direction in the explicit presence of matter and pre-matter should also be considered.

\subsection{The superbradyon hypothesis}
\vskip2mm
Superbradyons (superluminal preons) provide a simple explicit example of new ingredients that alternative cosmologies can naturally incorporate in pre-Big Bang scenarios. Again, the existence of a privileged rest frame for each comoving observer is naturally assumed \cite{Gonzalez-Mestres1996,Gonzalez-Mestres1995bis}. Superbradyons can be the basic constituents of the fundamental vacuum tacitly considered in the SST approach. 

In a limit where the usual kinematical concepts would still make sense for such objects, a simple choice for the relation between their energy ($E_s$), momentum ($p_s$) and velocity ($v_s$) would be \cite{Gonzalez-Mestres1995bis}:
\begin{eqnarray}
E_s~=~c_s~(p_s^2~+~m_s^2 ~c_s^2)^{1/2} \\
p_s~=~m_s~v_s~(1 ~-~v_s^2~c_s^{-2})^{-1/2}
\end{eqnarray}
where $m_s$ is the superbradyon mass and $c_s$ its critical speed assumed to be much larger than the speed of light $c$, just as $c$ is about a million times the speed of sound. 

Free superbradyons, if they exist, are usually assumed to have in most cases very weak direct interactions with laboratory conventional matter. The very small distance scales involved in superbradyon confinement inside standard particles, together with strong confinement forces, can be at the origin of such a property. When traveling at a speed larger than $c$, free superbradyons can spontaneously emit "Cherenkov" radiation in the form of standard particles \cite{Gonzalez-Mestres1995bis}. Remnant superbradyons may form in the present epoch a sea of particles with speeds close to $c$ that would be part of the cosmic dark matter \cite{Gonzalez-Mestres1996,Gonzalez-MestresCRIS2010}. 

If the ultimate constituents of matter can travel at a speed much faster than that of light and the vacuum can expand similarly, the very early Universe is expected to have naturally undergone a very fast expansion. Then, the horizon problem disappears and there is no longer any need for inflation \cite{Gonzalez-Mestres1997b,Gonzalez-Mestres1995bis}. The superbradyon hypothesis is just an illustration of the new physics that may be present in pre-Big Bang cosmologies.

Furthermore, in the case of the SST the expansion of the Universe basically follows an intrinsic geometric law ($H ~ = ~t^{-1}$), potentially perturbed by matter interactions \cite{Gonzalez-MestresCrete2013,Gonzalez-MestresSSTFriedmann}. This law is initially defined without any specific space variable, the cosmic time providing the only effective space scale \cite{Gonzalez-MestresPlanckSST,Gonzalez-MestresSSTFriedmann}. The comparison between the velocity of the Universe expansion and any critical speed of matter or pre-matter becomes possible only when matter and its constituents are explicitly introduced. Then, in the presence of explicit distance units, the Universe may turn out to be very large and to expand very quickly as compared to the critical speed of any form of matter or pre-matter.     

\subsection{The formation of standard matter}
\vskip2mm
The kind of pattern just described (pre-Big Bang and/or SST) provides natural alternatives to the standard cosmological mechanisms. In particular, Pre-Big Bang and SST approaches can naturally incorporate a very fast expansion of the early standard matter Universe. The formation of standard matter with its specific laws of Physics is expected to imply by itself an important phase transition.  

SST-based cosmologies imply a permanent expansion of the physical vacuum, possibly suggesting in our standard time language that the evolution of the Universe is driven by a fundamental instability. Then, in our standard matter Universe the effective vacuum structure and the basic parameters of the conventional laws of Physics can naturally be time-dependent. 

The standard cosmological constant and its usual phenomenological role 
are not required in this approach. We even do not necessarily expect the standard boson fields and harmonic-oscillator zero modes to be permanently condensed in vacuum in the absence of surrounding standard matter \cite{Gonzalez-MestresCrete1,Gonzalez-MestresIU}. The formation of conventional matter may just have been the emergence of vacuum excitations similar to phonons, solitons... \cite{Gonzalez-Mestres1997b,Gonzalez-Mestres1995bis} without really changing the basic (preonic ?) vacuum structure. It is even not obvious that our standard matter will be present in most of the available cosmic space. Instead, the SST cosmic space can be much larger even if, contrary to standard schemes, the global geometric curvature term from SST will play a leading role in the relevant modified Friedmann-like equations for the conventional matter Universe such as (2) \cite{Gonzalez-Mestres2014,Gonzalez-MestresCrete2013}. 

As standard matter will nucleate inside a pre-existing and already expanding universe with a pre-existing fundamental matter or pre-matter, fluctuations allowing for galaxy formation will be a natural phenomenon. A simple scenario would be to assume that standard matter is formed through many nucleation points associated to local type I phase transitions. Then, the existence of a local privileged space direction from SST can manifest itself leading to rotational modes around this direction for each nucleation center and, subsequently, to an associated CMB polarization incorporating $B$-modes.   

In such a scenario, latent heat can help to generate more standard matter. But the global expansion of the Universe is a pre-existing phenomenon led by the SST geometry, even if the energy released by a type I phase transition associated to the formation of standard matter is expected to be at the origin of local expansion effects for the conventional matter Universe. 

Contrary to the standard inflationary pattern, pre-Big Bang cosmologies do not need the Universe to be isotropic as seen by a comoving observer \cite{Gonzalez-MestresPrivSpDir,Gonzalez-MestresSSTFriedmann}. The spinorial space-time provides an explicit example of a different scenario \cite{Gonzalez-MestresCrete1,Gonzalez-MestresCRIS2010} using the original cosmic coordinates. 

As explained above, using a spinorial space-time appears as a natural choice, as the fermion wave functions do not correspond to representations of the real space rotation group SO(3) but of its covering group SU(2). We therefore expect the SST to be closer to the ultimate fundamental structure, and the early Universe (combining the Universe evolution with the internal particle structure) to be an appropriate framework to try to check its relevance. Instead of the standard Big Bang model with cosmic inflation, the observation of primordial CMB $B$-modes by BICEP2 may lead to a confirmation of the SST pattern. This would not be the only potential evidence for a cosmic SST, as Planck results already suggest \cite{PlanckAniso} that such a cosmic local space anisotropy can indeed be a real observable feature of our Universe. Obviously, further work is required.

The existence of a privileged rest frame for each comoving observer does not prevent standard relativity from remaining valid to a very good approximation in a wide range of energies and distance scales \cite{Gonzalez-MestresCrete2013bis,Gonzalez-MestresCrete2}. The situation is similar for phonons and solitons in a solid lattice, as long as parameters such as the lattice spacing can be neglected. But the possible existence of a PSD is a specific signature of the spinorial space-time and requires original investigation in cosmology and particle physics.

\subsection{Other pre-Big Bang cosmologies}
\vskip2mm 
Pre-Big Bang approaches can also produce primordial gravitational waves without any need for cosmic inflation. An explicit example was provided in \cite{Bogd1,Bogd2} with an initial singularity characterized by a gravitational instanton at $t$ = 0. Then, CMB $B$-modes can naturally result from primordial gravitational waves generated by the evolution of such an initial object and explicitly predicted by the authors.

Thus, primordial gravitational waves are not necessarily linked to an inflationary scenario. Similarly, the situation for vector perturbations can be substantially different from the standard approach. Not only in the case of SST-based cosmologies as explained above, but more generally. In standard cosmology, vector perturbations are ignored as a possible source of $B$-modes in the primordial cosmic microwave background polarization, arguing that they would be associated to vorticity in the plasma and quickly damped by inflation \cite{Takahashi}. But these considerations do not apply to most pre-Big Bang models where vector perturbations can be present already before Planck time, remain and develop in the absence of a standard inflationary evolution.  

More generally, pre-Big Bang models, including those based on the SST, can naturally avoid the basic problems of standard Big Bang cosmology that led to the development of the cosmic inflation scenario. Going beyond quantum mechanics eliminates intrinsic problems related to Planck scale. In particular, the global Universe can be much larger than the Planck distance at Planck time with no horizon problem, and contain a very large amount of energy. As low-energy symmetries do not necessarily become more exact at very high energy \cite{Gonzalez-MestresCRIS2010,Gonzalez-MestresUHECR}, the standard grand unification epoch is not necessarily present in its conventional form, and the monopole problem disappears. Similarly, Friedmann-like equations such as (2) automatically solve the flatness problem.

Pre-Big Bang scenarios remain a very open field where much work remains to be done with explicit formulations and tests, and that no astrophysical information invalidates at the present stage. The situation is similar concerning the possible ultimate constituents of matter, a crucial issue for pre-Big Bang patterns.

\section{The role of ultra-high energy physics}

Fundamental physics at ultra-high energy (UHE) remains by now poorly known \cite{Gonzalez-MestresCrete2013bis,Gonzalez-MestresCrete2}. 

It is even not yet clear \cite{AUGER1,AUGER2} if the observed fall of the ultra-high energy cosmic-ray (UHECR) spectrum is due to the Greisen-Zatsepin-Kuzmin (GZK) cutoff or corresponds to the maximum energies available at existing astrophysical sources. With such a limitation, it seems difficult to interpret data \cite{TA} on UHECR traveling on moderate extragalactic distances. Similarly, there is no real proof of the validity of models and algorithms used to describe UHECR interactions.

Exploring as far as possible the properties of UHECR with a search for signatures of new physics is an important task for a better understanding of the basic physics involved in the early Universe. Systematic tests of Lorentz symmetry at UHE \cite{Gonzalez-Mestres1996,Gonzalez-Mestres1997b} should be pursued and completed with tests of all fundamental principles of standard physics \cite{Gonzalez-MestresCRIS2010,Gonzalez-Mestres2009b}. The validity of all low-energy symmetries at very high energy also deserves a careful study \cite{Gonzalez-MestresCrete2,Gonzalez-MestresUHECR}. 

At $10^{19}$ eV, the proton conventional mass term is $\sim ~4.10^{-21}$ times its total kinetic energy. The relative (negative) contribution of this mass term to the proton velocity has a similar weight. Small departures from standard physics can therefore lead to detectable effects at high enough energy.

Another important issue, in connection with the subject dealt with here, is whether UHECR can be sensitive to the privileged space direction generated by the SST. In particular, possible correlations between high-energy cosmic rays and the recent Planck observation on the CMB anisotropy \cite{PlanckAniso} should be systematically explored through a long-term program.

Accelerator experiments can also contribute to this search for new physics, not only concerning possible supersymmetric patterns and superbradyon production, but also for approaches that escape the standard theorems based on the Poincar\'{e} group \cite{Gonzalez-MestresCrete1,Gonzalez-MestresPreBB1}. An example can be the search for elementary particles with spin 3/2, 2 and higher potentially generated in a SST approach as previously described. Such particles can also exist as high-energy cosmic rays resulting from the decay of heavy objects (e.g. superbradyons), as already suggested in \cite{Gonzalez-Mestres1996,Gonzalez-Mestres1996SBdec} for UHECR.

\section{Conclusion and comments}

There is by now no serious reason to consider the possible existence of CMB B-modes of primordial origin, suggested by BICEP2 data, as a potentially compelling evidence for the standard Big Bang and cosmic inflation. 

Instead, such a result can be one of the natural signatures of a pre-Big Bang era or, simply, of a space-time geometry (the spinorial space-time) best adapted than the conventional real space-time to the existence of fermions among the elementary particles of standard Physics.  

Planck results on CMB anisotropy with a possible local privileged space direction \cite{PlanckAniso} already suggest that nonstandard cosmological phenomena may be at work. If confirmed, and together with Planck data, the recent BICEP2 result can mark the emergence of unprecedented evidence for physics beyond the standard model, with a new space-time geometry beyond conventional relativity and leading to a new cosmology that would replace the pattern based on $\Lambda $CDM and inflation.

The observed acceleration of the expansion of the standard matter Universe can also be explained by this new cosmology \cite{Gonzalez-MestresCrete2013,Gonzalez-MestresCrete1}. The possibility that in pre-Big Bang and SST-like patterns the just nucleated standard matter reacts (including gravitationally) to the pre-existing fast expansion of the Universe has been explicitly considered in previous papers. See, for instance, the conclusion of \cite{Gonzalez-MestresPlanckSST}. Then, the equation $H ~ t ~= ~1$ can be a natural asymptotic limit at large $t$ as the matter density decreases.

As the considered pre-Big Bang models already incorporate a phase transition associated to the formation of standard matter, cosmic inflation is not necessary and may even appear as an artificial trick as compared to matter nucleation from ultimate constituents. Similarly, pre-Big Bang scenarios can naturally generate primordial vector perturbations producing CMB $B$-modes and gravitational waves leading to the same effect. The theoretical uncertainty in the interpretation of possible primordial $B$-modes of the CMB polarization has clearly been underestimated in recent statements.

A more detailed discussion of these important issues and of the ideas dealt with in this paper will be presented elsewhere.

\end{document}